\documentclass[onecolumn,preprintnumbers,amsmath,amssymb]{revtex4}
\usepackage{array}
\usepackage{booktabs}
\usepackage{tabu}
\usepackage{dcolumn}
\usepackage{amsmath}
\usepackage{amsfonts}
\usepackage{amssymb}
\usepackage{graphicx}
\usepackage{subfigure}
\usepackage{graphicx}% Include figure files
\usepackage{dcolumn}% Align table columns on decimal point
\usepackage{bm}% bold math
\usepackage{xcolor}
\usepackage{amsthm}
\def\be{\begin{equation}}
\def\ee{\end{equation}}
\def\bea{\begin{eqnarray}}
\def\eea{\end{eqnarray}}
\def\f{\frac}
\def\n{\nonumber}
\def\l{\label}
\def\p{\phi}
\def\o{\over}
\def\R{\rho}
\def\pa{\partial}
\def\om{\omega}
\def\na{\nabla}
\def\P{\Phi}

\begin{document}

\title{Gravitationally-Induced Conversion of Local Coherence to Entanglement}% Force line breaks with \\

\author{H. Dolatkhah}
\email{h.dolatkhah@gmail.com}
\affiliation{Department of Physics Education, Farhangian University, P.O. Box 14665-889, Tehran, Iran}
\author{S. Salimi}
\affiliation{Department of Physics, University of Kurdistan, P.O.Box 66177-15175, Sanandaj, Iran}
\author{S. Haseli}
\affiliation{Faculty of Physics, Urmia University of Technology, Urmia, Iran.\\}

\date{\today}% It is always \today, today,

%%%%%%%%%%%%%%%%%%%%%
%%%%%%%%%%%%%%%%%%%%
%%%%%%%%%%%%%%%%%%%%%%
%%%%%%%%%%%%%%%%%%%%%%%%

\def\be{\begin{equation}}
  \def\ee{\end{equation}}
\def\bea{\begin{eqnarray}}
\def\eea{\end{eqnarray}}
\def\f{\frac}
\def\n{\nonumber}
\def\l{\label}
\def\p{\phi}
\def\o{\over}
\def\R{\rho}
\def\pa{\partial}
\def\om{\omega}
\def\na{\nabla}
\def\P{\Phi}
%\nofiles

%=============================================================%
%=============================================================%
%============== Abstract =======================================%
%=============================================================%
%=============================================================%
\begin{abstract}
In recent years, the quantum nature of gravity has attracted significant attention as one of the most important problems in modern physics. Here, we analyze the mechanism of gravitationally induced entanglement from the perspective of quantum resource theory. Building on the interferometric framework of Bose \textit{et al.} [\href{PHYSICAL REVIEW LETTERS volume 119, Article number: 240401 (2017)}{Phys. Rev. Lett. 119, 240401 (2017)}], we show that the gravitational interaction acts as a unitary channel, redistributing quantum resources between two spatially superposed masses. Specifically, we demonstrate that the resulting bipartite entanglement originates from the coherent conversion of local quantum coherence—initially present in each subsystem—into shared nonlocal correlations. We derive exact, analytical complementarity relations quantifying this conversion, link the decay of local coherence directly to the growth of entanglement, and support these findings with numerical simulations. Our results clarify the underlying mechanism and establish gravity as a coherence-to-entanglement conversion channel, offering a refined interpretive basis for forthcoming experimental tests. Crucially, we show that initial coherence is a necessary condition for entanglement generation and that its degree bounds the maximum achievable entanglement, with maximal entanglement requiring initial maximal coherence.
\end{abstract}
%\pacs{04.50}
%\keywords{keyword.}%Use showkeys class option if keyword
\maketitle

%%%%%%%%%%%%%%%%%%%%%%%%%%%%%%%%%%%%%%%%%%%%%%%%%%%%%%%%%%%%%%%%%%%%%%%%%%%%
%%%%%%%%%%%%%%%%%%%%%%%%%%%%%%%%%%%%%%%%%%%%%%%%%%%%%%%%%%%%%%%%%%%%%%%%%%%%
%%%%%%%%%%%%%%%%%%%%%%%%%%%%%%%%%%%%%%%%%%%%%%%%%%%%%%%%%%%%%%%%%%%%%%%%%%%%
%%%%%%%%%%%%%%%%%%%%%%%%%%%%%%%%%%%%%%%%%%%%%%%%%%%%%%%%%%%%%%%%%%%%%%%%%%%%
%============  Sec.I (Introduction)  =======================================
%%%%%%%%%%%%%%%%%%%%%%%%%%%%%%%%%%%%%%%%%%%%%%%%%%%%%%%%%%%%%%%%%%%%%%%%%%%%
%%%%%%%%%%%%%%%%%%%%%%%%%%%%%%%%%%%%%%%%%%%%%%%%%%%%%%%%%%%%%%%%%%%%%%%%%%%%
%%%%%%%%%%%%%%%%%%%%%%%%%%%%%%%%%%%%%%%%%%%%%%%%%%%%%%%%%%%%%%%%%%%%%%%%%%%%
%%%%%%%%%%%%%%%%%%%%%%%%%%%%%%%%%%%%%%%%%%%%%%%%%%%%%%%%%%%%%%%%%%%%%%%%%%%%
\section{Introduction}	%) A SECTION HEADING
One of the most important questions in modern physics is whether gravity possesses an intrinsic quantum nature. So far, many efforts have been made to investigate and answer this question \cite{Bose,Marletto,Kafri,Bose1,Marletto1,Abrahao,Sharifian,Sugiyama,Gallock,Christodoulou,Christodoulou1,Haine,Carlesso,Pedernales,Matsumura,Perche,Perche2,Hall,Howl,Rijavec,Sugiyama2,Wan,van,Wood,Toros,Gunnink,Hidaka,Li}. A central strategy toward addressing this question is to determine whether gravity can mediate entanglement between two massive objects with embedded spins\cite{Bose,Marletto}. In particular, the proposal introduced by Bose \textit{et al.} demonstrates that if two systems become entangled solely through their mutual gravitational interaction, then the mediator cannot be classical, since classical interactions constrained to local operations and classical communication (LOCC) cannot generate entanglement \cite{Bose}. While this criterion establishes a necessary condition for the quantumness of gravity, it does not explain the mechanism by which entanglement is generated. In particular, the origin, redistribution, and flow of quantum resources under gravitational evolution are only implicitly treated. In this work, we address this conceptual gap by examining the gravitational interaction, specifically the model presented by Bose \textit{et al.} \cite{Bose}, from the perspective of quantum resource theory.
We demonstrate that the gravitational interaction constitutes a unitary, coherence-preserving operation. Consequently, the quantum coherence of the total system remains constant under gravitational transformation. Moreover, we show that the gravitational interaction acts as a coherent resource converter, dynamically redistributing local quantum coherence—initially stored in the spatial superposition of each mass—into shared bipartite entanglement. The degree of generated entanglement is quantitatively determined by, and bounded by, the initial coherence. This leads to two fundamental constraints: (i) initial coherence is a necessary resource for entanglement generation (vanishing coherence implies zero entanglement), and (ii) maximal entanglement is attainable only if the subsystems are initially maximally coherent.\\
Furthermore, we derive exact complementarity relations that quantify this coherence-entanglement conversion. Although related complementarity has been discussed in other contexts \cite{Streltsov2,Xi,Pan}, including gravitational settings \cite{Maleki,Chen}, our work takes a different approach by identifying gravity as a coherence-preserving unitary transformation and by deriving exact analytical complementarity relations from an explicit gravitationally induced dynamical model.\\
The paper is organized as follows. In Sec. \ref{Sec2}, the model in which two neutral test
masses $A$ and $B$, with masses $m_{A}$ and $m_{B}$, respectively, become entangled through gravitational interaction will be  briefly explained. Sec. \ref{Sec3} provides a concise review of the quantum resource theory of coherence. In Sec. \ref{Sec4}, the influence of the gravitational interaction on the quantum coherence is analyzed. Finally, the last section is dedicated to conclusions and discussion.

%%%%%%%%%%%%%%%%%%%%%%%%%%
%%%%%%%%%%%%%%%%%%%%%%%%%%%
%%%%%%%%%%%%%%%%%%%%%%%%%%
%%%%%%%%%%%%%%%%%%%%%%%%%%
\section{Gravitational interaction between two masses}\label{Sec2}
%%%%%%%%%%%%%%%%%%%%%%%%%%%
%%%%%%%%%%%%%%%%%%%%%%%%%%%

To clarify how the states of two neutral test masses  $m_{A}$ and $m_{B}$​ become entangled via gravitational interaction, a summary of the theory presented in \cite{Bose} is provided. Consider two spatially delocalized two-level systems, $A$ and $B$, as shown in Fig. \ref{fig1}. Each system can occupy one of two spatially separated locations denoted by $|L\rangle$ and $|R\rangle$. Imagine the centers of $|L\rangle$ and $|R\rangle$  are separated by a distance  $\Delta x$, here we suppose that $\Delta x=\Delta x_{A}=\Delta x_{B}$. These states are described by localized Gaussian wavepackets with widths  $\Delta x$ so that one can ensure $\langle L|R\rangle =0$, ensuring the two states are effectively orthogonal. Additionally, the centers of the superpositions of the two masses are separated by a distance $d$. This distance is chosen to be large enough so that even at their closest approach  $(d-\Delta x)$, one can neglect short-range interactions such as the Casimir-Polder force. Due to the spatial separation, different branches of the spatial superposition experience different gravitational interaction energies, as the distances between the masses vary in each configuration. This results in different phase evolution rates. Under these conditions, the time evolution of the joint quantum state of the two masses is governed solely by their mutual gravitational interaction. Initially, each mass is prepared in a superposition state:
%%%%%%%%%%%%%%%%%%%%%%%%%%%
%%%%%%%%%%%%%%%%%%%%%%%%%%%%%%%%%%%
%%%%%%%%%%%%%%%%%%%%%%%%%%%%%%%%%%%%
\begin{figure}[ht] 
\centering
\includegraphics[width=8cm]{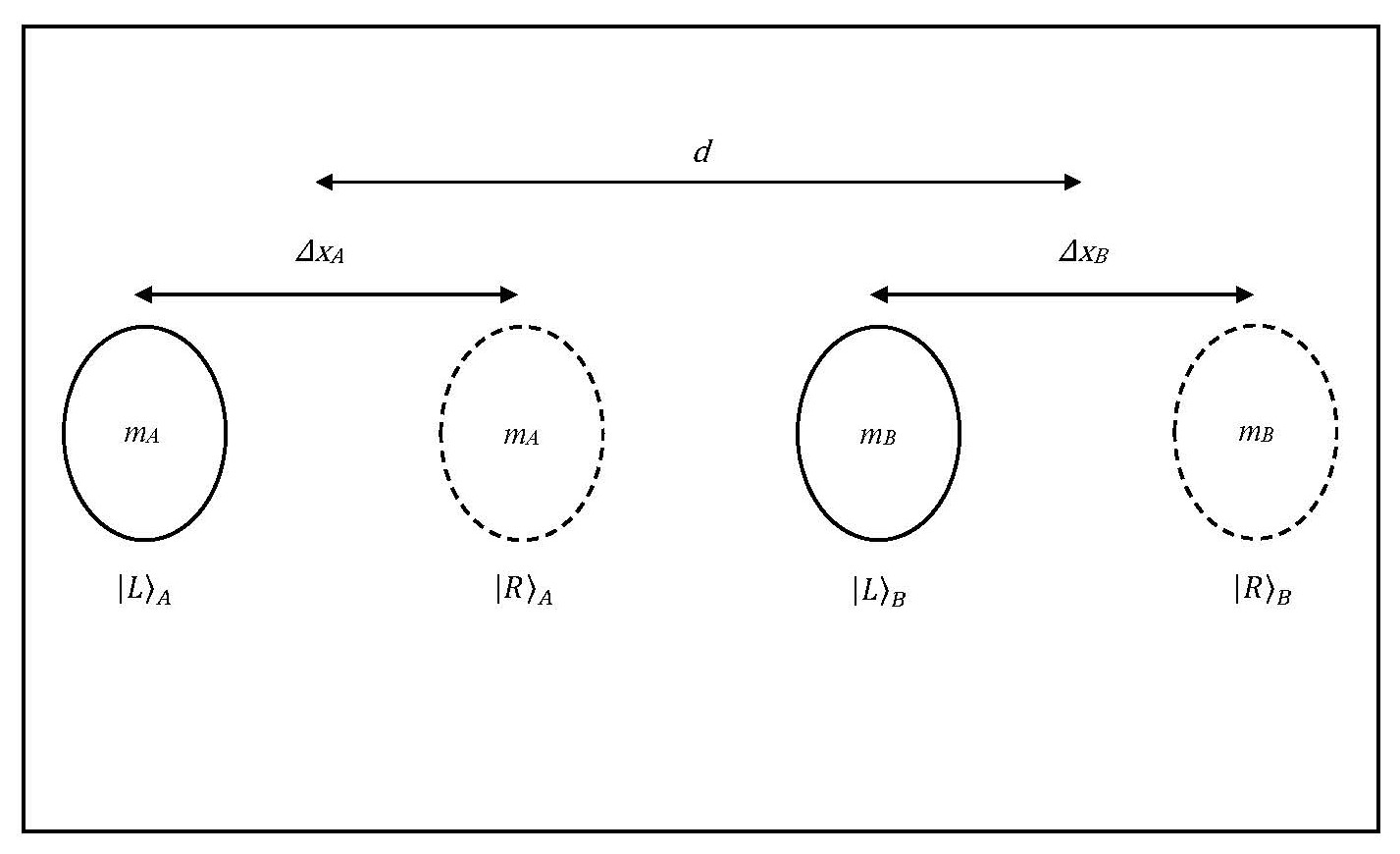}
\caption{Two test masses are maintained adjacent to each other in a superposition of spatially localized states $|L\rangle$ and $|R\rangle$.}\label{fig1}
\end{figure}
 
%%%%%%%%%%%%%%%%%%%%%%%%%%
%%%%%%%%%%%%%%%%%%%%%%%%%%%%%%%
\begin{eqnarray}
|\psi(0)\rangle_{A} = \frac{1}{\sqrt{2}} (|L\rangle_A + |R\rangle_A), \\ |\psi(0)\rangle_{B}=\frac{1}{\sqrt{2}}(|L\rangle_B + |R\rangle_B). 
\end{eqnarray}
Therefore, the initial joint state of the two masses is
\begin{equation}\label{intial state}
|\psi(0)\rangle_{AB} = \frac{1}{\sqrt{2}} (|L\rangle_A + |R\rangle_A) \otimes \frac{1}{\sqrt{2}}(|L\rangle_B + |R\rangle_B). 
\end{equation}
Following the model of Bose \textit{et al.} \cite{Bose}, the state after a gravitational interaction time $\tau$ evolves to
\begin{equation}\label{evolved state}
|\psi(\tau)\rangle_{AB} = \frac{e^{i\phi}}{2} (|LL\rangle_{AB} + e^{i\Delta\phi_{LR}}|LR\rangle_{AB} + e^{i\Delta\phi_{RL}}|RL\rangle_{AB} + |RR\rangle_{AB}), 
\end{equation}
in which $\Delta\phi_{RL}=\phi_{RL}-\phi, \Delta\phi_{LR}=\phi_{LR}-\phi $ and
\begin{equation}
 \phi_{RL}\sim\frac{Gm_{1}m_{2}\tau}{h(d-\Delta x)}, \quad  \phi_{LR}\sim\frac{Gm_{1}m_{2}\tau}{h(d+\Delta x)}, \quad
  \phi\sim\frac{Gm_{1}m_{2}\tau}{hd}.
\end{equation}
Each mass can now be considered as an effective "orbital qubit," where its two states correspond to the spatial states $|L\rangle$ and $|R\rangle$—referred to as orbital states \cite{Bose}. This simplifies the problem to a two-qubit system, where the spatial degree of freedom physically encodes the qubit states. For entanglement to exist between the qubits, the total state must be inseparable. This means that the following states must not be identical,
$\frac{1}{\sqrt{2}}(|L\rangle_{B} + e^{i\Delta\phi_{LR}}|R\rangle_{B})$ and $\frac{1}{\sqrt{2}}(e^{i\Delta\phi_{RL}}|L\rangle_{B} + |R\rangle_{B})$ 
, this happens when $ \Delta\phi_{LR}+\Delta\phi_{RL}\neq 2n\pi$ in which $n$ is an integer. In this case the total state cannot be factorized and is thereby an entangled state of the two orbital qubits.
%%%%%%%%%%%%%%%%%%%%%%%%%%%%%%%%%%

\section{Quantum resource theory of coherence}\label{Sec3}
%%%%%%%%%%%%%%%%%%%%%%%%%%%%%%
 Quantum coherence is one of the most important features of quantum physics and it is a prominent physical resource in many quantum information\cite{Streltsov}. In this section we briefly introduce the resource theory of coherence. The quantum resource theory is defined by determining the set of free states and free operations. One can consider  the incoherent states and the incoherent operations as free states and free operations, respectively, which define as follows:\\
\textbf{Definition 1 (Incoherent states).} A fixed orthonormal basis $\mathcal{B} :=  \lbrace \vert i \rangle \rbrace^{d-1}_{i=0} $ is chosen. A quantum state is called incoherent if its density matrix is diagonal in this basis, i.e., if it can be written as
\begin{equation}
\hat{\delta} =\sum_{i=0}^{d-1} p_{i}\vert i \rangle\langle i \vert,
\end{equation}
where $p_{i}\geqslant 0$ and $\sum_{i} p_{i}=1$.  This set of quantum states are displayed by $\mathcal{I}$ and $\hat{\delta}\in \mathcal{I}$.\\
\textbf{Definition 2 (Incoherent operations).} An incoherent operation has a Kraus representation such that  $K_{m} \mathcal{I} K_{m}^{\dag} \subseteq \mathcal{I}$ for all $m$. It is clear that incoherent operations mapping incoherent states onto incoherent states.\\
 To quantify coherence, several measures have been proposed \cite{Streltsov}. The $l_1$-norm of coherence and relative entropy of coherence \cite{Baumgratz} can be considered as two proper coherence measures. The $l_1$-norm of coherence is defined as
\begin{equation}
C_{l_{1}}(\rho)=\sum_{i\neq j}\vert\rho_{ij}\vert,
\end{equation}
which corresponds to the sum of the absolute values of all off-diagonal elements of the density matrix in the chosen basis. The relative entropy of coherence is defined as

\begin{equation}\label{relative coherence}
C_{r}(\rho)=S(\rho^{diag})-S(\rho),
\end{equation}
where $S(\rho)=- Tr(\rho \log_{2}\rho)$ is the von Neumann entropy and $\rho^{diag}$ is obtained from removing all off-diagonal elements of the state density matrix.\\
We now introduce several key definitions from coherence resource theory that are central to our analysis.\\
\textbf{Definition 3 (Maximally coherent states).} A state $|\psi \rangle$ is called maximally coherent with respect to a fixed reference basis  $\lbrace\vert j \rangle\rbrace$ if it can be written in the form \cite{Peng}
\begin{equation}\label{MCS}
|\psi\rangle = \frac{1}{\sqrt{d}} \sum_{j=0}^{d-1} e^{i\theta_j} |j\rangle,
\end{equation}
where $\theta_{j}\in [ 0,2\pi)$ are arbitrary phases. The set of all such states is denoted by $\mathcal{S}_{MCS}$.\\
\textbf{Definition 4 (Unitary incoherent operations).} It was proven in Ref. \cite{Peng} that the unitary incoherent operations have the following form
\begin{equation}
 U=\sum_{j}e^{i\theta_{j}}|\alpha(j) \rangle\langle j|,
 \end{equation}
in which ${\alpha(j)}$ is a relabeling of ${j}$.
It is also shown an incoherent unitary operation of the bipartite quantum system is of the form \cite{Kraft}
 \begin{equation}\label{bui}
 U=\sum_{i,j}e^{i\varphi_{ij}}|\Pi(ij) \rangle\langle ij|,
 \end{equation}
in which $\Pi$ is a permutation of the pairs $(i,j)$.\\ 
\textbf{Definition 5 (Coherence-preserving operations).} An operation is coherence preserving if and only if it is unitary and incoherent \cite{Peng}.\\
%%%%%%%%%%%%%%%%%%%%%%%%%%%%%%%   
%%%%%%%%%%%%%%%%%%%%%%%%%
%%%%%%%%%%%%%%%%%%%%%%%%%%%%%%%%%%%%
\begin{figure}[ht] 
\centering
\includegraphics[width=6cm]{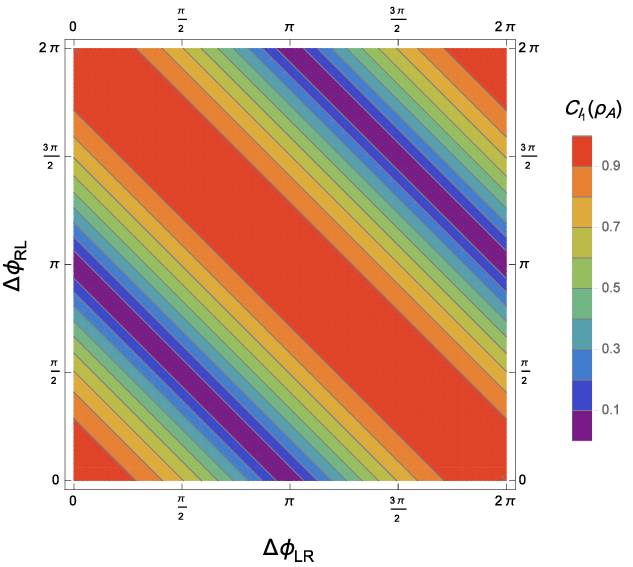}
\caption{The $l_1$-norm of coherence $C_{l_{1}}(\rho_{A})$ plotted as a function of the phases $\Delta \phi_{L R}$ and $\Delta \phi_{R L}$.}\label{fig2}
\end{figure}
%%%%%%%%%%%%%%%%%%%%%%%%%%%%%%%%%
%%%%%%%%%%%%%%%%%%%%%%%%%%%%%%%%%%%%
\begin{figure}[ht] 
\centering
\includegraphics[width=6cm]{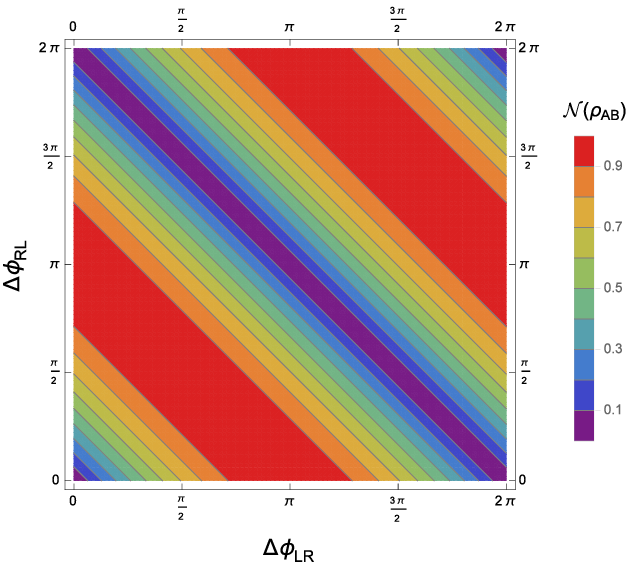}
\caption{The negativity $\mathcal{N}(\rho_{AB})$ plotted as a function of the phases $\Delta \phi_{L R}$ and $\Delta \phi_{R L}$.}\label{fig3}
\end{figure}
%%%%%%%%%%%%%%%%%%%%%%%%%%%%%%%%%

%%%%%%%%%%%%%%%%%%%%%%%%%%%%%%%%%%
%%%%%%%%%%%%%%%%%%%%%%%%%%%%%%%%%
\section{influence of the gravitational interaction on the quantum coherence}\label{Sec4}
%%%%%%%%%%%%%%%%%%%%%%%%%%%%%%%%%%%%
\begin{figure}[ht] 
\centering
\includegraphics[width=6cm]{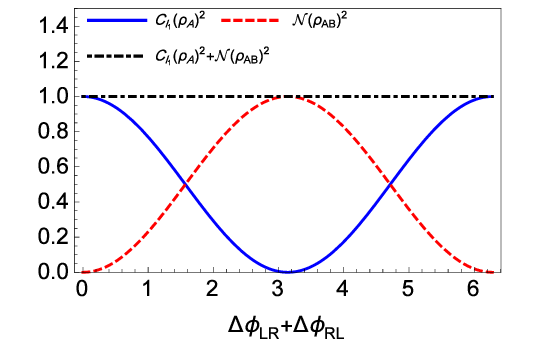}
\caption{The squared $l_1$-norm of coherence and squared negativity as functions of the total accumulated phase $\Delta \phi_{L R}+\Delta \phi_{R L}$. The blue solid line represents $C_{l_{1}}^{2}(\rho_{A})$, the red dashed curve represents $\mathcal{N}^{2}(\rho_{AB})$, and the black dot-dashed curve represents their sum  $C_{l_{1}}(\rho_{A})^{2}+\mathcal{N} (\rho_{AB})^{2}$.}\label{fig4}
\end{figure}
%%%%%%%%%%%%%%%%%%%%%%%%%%%%%%%%%
%%%%%%%%%%%%%%%%%%%%%%%%%%%%%%%%%%%%%
%%%%%%%%%%%%%%%%%%%%%%%%%%%%%%%%%%%%%
%%%%%%%%%%%%%%%%%%%%%%%%%%%%%%%%%
%%%%%%%%%%%%%%%%%%%%%%%%%%%%%%%%%
%%%%%%%%%%%%%%%%%%%%%%%%%%%%%%%%%%%%%%
%%%%%%%%%%%%%%%%%%%%%%%%%%%%%%%%%%%%%%
\begin{figure}[ht] 
\centering
\includegraphics[width=6cm]{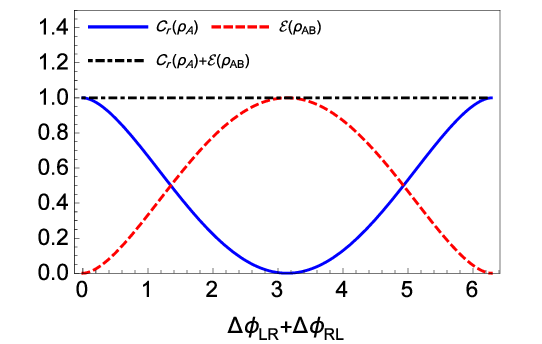}
\caption{The relative entropy of coherence and the von Neumann entropy of the reduced state of the subsystem $A$ as functions of the total accumulated phase $\Delta \phi_{L R}+\Delta \phi_{R L}$. The blue solid line represents $C_{r}(\rho_{A})$, the red dashed curve represents $\mathcal{E}(\rho_{AB})$, and the black dot-dashed curve represents their sum  $C_{r}(\rho_{A})+\mathcal{E}(\rho_{AB})$.}\label{fig5}
\end{figure}
\begin{figure}[ht] 
\centering
\includegraphics[width=6cm]  {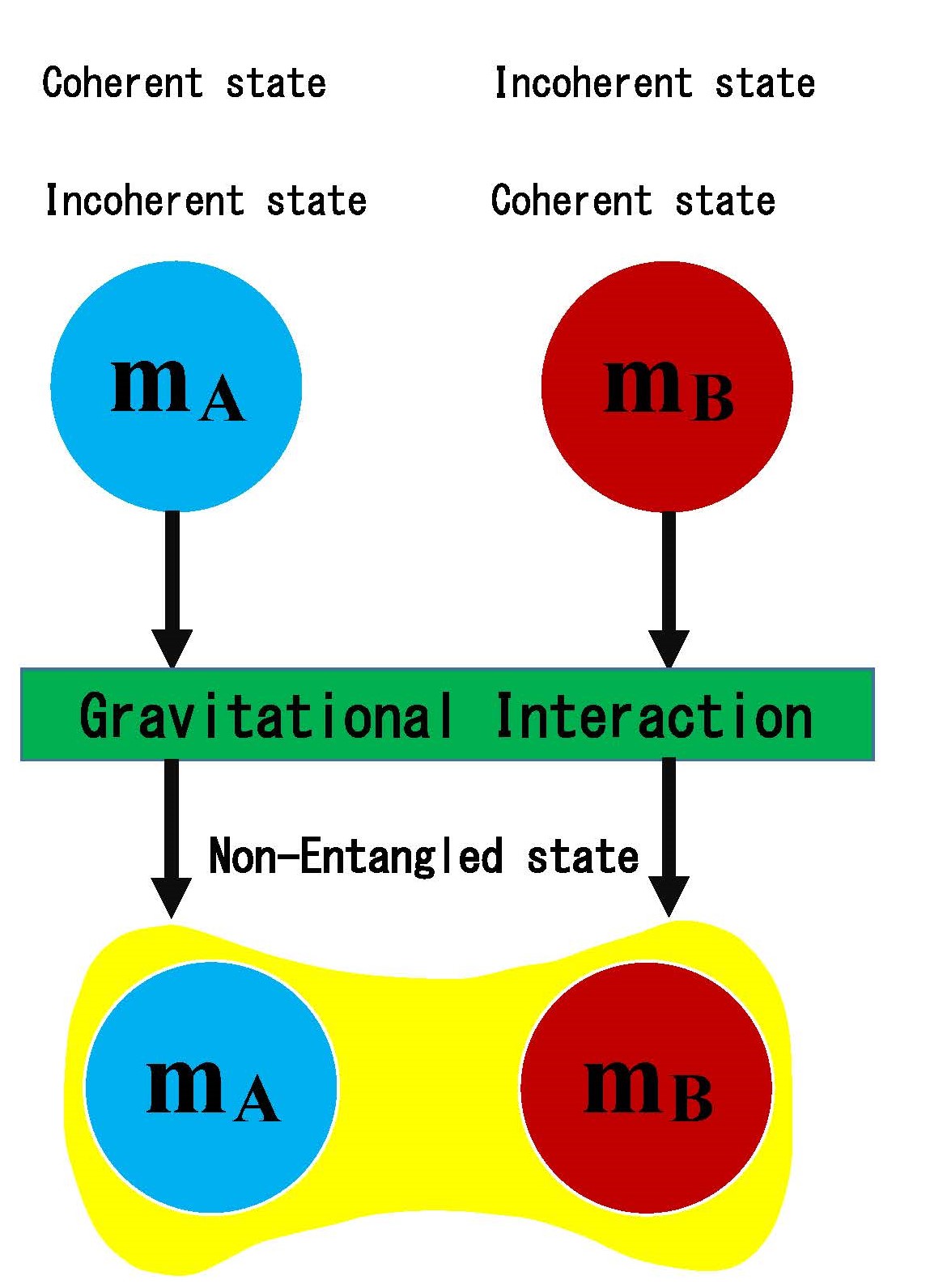}
\caption{ For the generation of entanglement under gravity, the existence of initial coherence in the subsystems is necessary.}\label{fig6}
\end{figure}
\begin{figure}[ht] 
\centering
\includegraphics[width=6cm]  {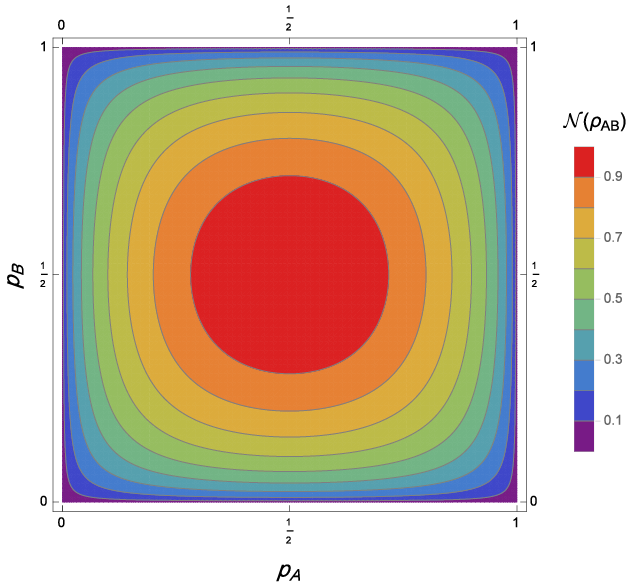}
\caption{ The negativity $\mathcal{N}(\rho_{AB})$ as a function of the $p_{A}$ and $p_{B}$, where $\Delta\phi_{LR} + \Delta\phi_{RL}=\pi$.}\label{fig7}
\end{figure}

%%%%%%%%%%%%%%%%%%%%%%%%%%%%%%%%%%%%%%%
%%%%%%%%%%%%%%%%%%%%%%%%%%%%%%%%%%%%%%%
%%%%%%%%%%%%%%%%%%%%%%%%%%%%%%%%%%%%%%
Now, let us study the influence of the gravitational interaction on the quantum coherence. 
The density operator for a pure state $ |\psi\rangle $ is defined as  $\rho=|\psi\rangle\langle\psi| $. For the initial state in  Eq.\;(\ref{intial state}) one has
 \begin{equation}\label{M initial state}
\rho_{AB}(t=0)=\frac{1}{4}\left(\begin{array}{llll}
1 & 1 & 1 & 1 \\
1 & 1 & 1 & 1 \\
1 & 1 & 1 & 1 \\
1 & 1 & 1 & 1
\end{array}\right),
\end{equation}  
in the basis ${\vert L L\rangle,|L R\rangle,|R L\rangle, |R R\rangle}$. According to Eq.\;(\ref{MCS}), it is easy to see that this state is a maximally coherent state. Also we know that this state is a product state, which means it has no correlation between the subsystems. After the evolution time $\tau$, the density matrix becomes
\begin{equation}\label{M evolved state}
\rho_{AB}(t=\tau)=\frac{1}{4}\left(\begin{array}{cccc}
1 & e^{-i \Delta \phi_{L R}} & e^{-i \Delta \phi_{R L}} & 1 \\
e^{i \Delta \phi_{L R}} & 1 & e^{\left.-i (\Delta \phi_{R L}-\Delta \phi_{L R}\right)} & e^{i \Delta \phi_{L R}} \\
e^{i \Delta \phi_{R L}} & e^{i\left(\Delta \phi_{R L}-\Delta \phi_{L R}\right)} & 1 & e^{i \Delta \phi_{L R}} \\
1 & e^{-i \Delta \phi_{R L}} & e^{-i \Delta \phi_{L R}} & 1
\end{array}\right).
\end{equation}
This state is also a maximally coherent states with this difference that based on entanglement theory, one can easily know that the state can be also a maximally entangled state for some special phases. Comparing these two states (Eqs.\;(\ref{M initial state}) and (\ref{M evolved state})), implies that though these two states are both the maximally coherent states, their reduced states and also the origin of their coherences can be completely different. To find the reason, consider the gravitational interaction discussed in section \ref{Sec2}, one can readily obtain its form as follows:
 \begin{equation}
 U_{G}=\sum_{i,j}e^{i\phi_{ij}}|ij \rangle\langle ij|,
 \end{equation}
with $i,j\in\lbrace L,R\rbrace$. Comparing this expression with the general form of a bipartite incoherent unitary in Eq.\;(\ref{bui}) shows that $U_{G}$ is an incoherent unitary operation. Therefore, the gravitational interaction introduced by Bose \textit{et al.} \cite{Bose} is coherence-preserving, which explains the conservation of total quantum coherence during the evolution.\\ Now, to clarify the gravitational action, we analyze the interplay between coherence and entanglement during the evolution. We proceed by examining two distinct cases, each employing different resource measures.\\
\noindent\textbf{Case I} \textbf{($l_1$-norm and negativity)}. The measures of coherence and entanglement are taken to be the $l_1$-norm of coherence and the negativity \cite{Peres,Horodecki}, respectively. The negativity $\mathcal{N}(\rho_{AB})$ can be defined by \cite{Horodecki2}
\begin{equation}
\mathcal{N}(\rho_{AB})=\parallel \rho_{AB}^{T_{A}}\parallel_{1}-1,
\end{equation}
where $\parallel .\parallel_{1}$ denotes the trace norm and $\rho^{T_{A}}$ represents the partial transposition of $\rho_{AB}$ with respect to subsystem $A$.
\\ 
By tracing out subsystem $B$,  the density matrix for mass $A$ can be expressed as
\begin{equation}
\rho_A=\operatorname{Tr}_B(\rho_{AB}(t=\tau))=\frac{1}{2}\left(\begin{array}{cc}
1 &e^{i\left({\frac{\Delta \phi_{L R}-\Delta \phi_{R L}}{2}}\right)} \cos \left(\frac{\Delta \phi_{L R}+\Delta \phi_{R L}}{2}\right) \\
e^{-i\left({\frac{\Delta \phi_{L R}-\Delta \phi_{R L}}{2}}\right)}\cos \left(\frac{\Delta \phi_{L R}+\Delta \phi_{R L}}{2}\right) & 1
\end{array}\right).
\end{equation}
The $l_1$-norm of coherence for the above density matrix is 
\begin{equation}
C_{l_{1}}(\rho_{A})=\mid\cos \left(\frac{\Delta \phi_{L R}+\Delta \phi_{R L}}{2}\right)\mid,
\end{equation}
and one can obtain that the negativity is   
\begin{equation}
\mathcal{N}(\rho_{AB})=\mid \sin(\frac{\Delta\phi_{LR} + \Delta\phi_{RL}}{2})\mid.
\end{equation}
In Figures \ref{fig2} and \ref{fig3}, the $l_1$-norm of coherence and negativity are plotted as functions of the accumulated phases, respectively. The plots demonstrate an inverse correlation between the two quantities: the peak of negativity coincides with the zero point of $l_1$-norm of coherence, and conversely.\\ Using the above equations, one arrives at 
\begin{equation}\label{ComN}
C_{l_{1}}^{2}(\rho_{A})+\mathcal{N}^{2}(\rho_{AB})=1,
\end{equation}
which is a complementary relation. It is worth mentioning that in the case of the state in Eq.\;(\ref{M evolved state}) the negativity is equal to the concurrence \cite{Miranowicz}. Thus, one can rewrite the above equation as
 \begin{equation}\label{ComC}
C_{l_{1}}^{2}(\rho_{A})+\mathcal{C}^{2}(\rho_{AB})=1,
\end{equation}
where $\mathcal{C}(\rho_{AB})$ is the concurrence. These equalities (Eqs.\;(\ref{ComN}) and(\ref{ComC})) demonstrate that the sum of the squares of local coherence and bipartite entanglement is conserved, always equal to unity. Consequently, any increase in entanglement must be accompanied by a decrease in local coherence, and vice versa (see Fig. \ref{fig4}). \\ 
\noindent\textbf{Case II} \textbf{(Relative entropy of coherence and entanglement entropy)}. In this case, the measure of coherence is taken to be the relative entropy of coherence, while entanglement is measured by the von Neumann entropy of the reduced state of subsystem $A$, i.e., the entanglement entropy $\mathcal{E}(\rho_{AB})=S(\rho_{A})$. Regarding these measures, the coherence and entanglement are given respectively by 

\begin{equation}\label{relative coherence}
C_{r}(\rho_{A})=1-H_{2}(\dfrac{1+c}{2}).
\end{equation}
and 
\begin{equation}
\mathcal{E}(\rho_{AB})=S(\rho_{A})=H_{2}(\dfrac{1+c}{2}),
\end{equation}
where $H_{2}(c) = -c \log_{2} c - (1- c) \log_{2}(1- c)$ denotes the binary Shannon entropy function and $c=\cos \left(\frac{\Delta \phi_{L R}+\Delta \phi_{R L}}{2}\right)$. It is easy to note that
\begin{equation}\label{ComR}
C_{r}(\rho_{A})+\mathcal{E}(\rho_{AB})=1.
\end{equation}
This also establishes a complementary relation between entanglement and coherence. The evolution of coherence and entanglement in this scenario is illustrated in Fig. \ref{fig5}. Based on the results obtained so far, it can be concluded that gravity acts as a channel that converts the coherence of subsystems into entanglement, and vice versa.\\
%%%%%%%%%%%%%%%%%%%%%%%%%%%%%%%%%%%
%%%%%%%%%%%%%%%%%%%%%%%%%%%%%%%%%%%%
 %%%%%%%%%%%%%%%%%%%%%%%%%%%%%%%%%%
 %%%%%%%%%%%%%%%%%%%%%%%%%%%%%%%%%%%%%%%
 %%%%%%%%%%%%%%%%%%%%%%%%%%%%%%%%%%
 %%%%%%%%%%%%%%%%%%%%%%%%%%%%%%%%%%%
We now investigate how the initial coherence influences the entanglement generated by gravity. For this purpose, we generalize the initial state considered in Sec. \ref{Sec2} to an arbitrary pure product state and apply the same analysis. The initial state is taken as
 \begin{equation}
|\psi(0)\rangle_{AB} = (\sqrt{p_{A}}|L\rangle_A + \sqrt{1-p_{A}}|R\rangle_A) \otimes (\sqrt{p_{B}}|L\rangle_B +\sqrt{1-p_{B}} |R\rangle_B), 
\end{equation}
 where $0\leqslant p_{A(B)}\leqslant1$. The density operator for this initial state is
 \begin{equation}
\rho_{AB}(t=0)=\left(\begin{array}{llll}
|M_{LL}|^{2} & M_{LL}M^{*}_{LR} & M_{LL}M^{*}_{RL} & M_{LL}M^{*}_{RR} \\
M_{LR}M^{*}_{LL} & |M_{LR}|^{2} & M_{LR}M^{*}_{RL} & M_{LR}M^{*}_{RR}\\
M_{RL}M^{*}_{LL} & M_{RL}M^{*}_{LR} & |M_{RL}|^{2} & M_{RL}M^{*}_{RR} \\
M_{RR}M^{*}_{LL} & M_{RR}M^{*}_{LR} & M_{RR}M^{*}_{RL} & |M_{RR}|^{2}
\end{array}\right),
\end{equation} 
where $M_{LL}=\sqrt{p_{A}p_{B}}$, $M_{LR}=\sqrt{p_{A}(1-p_{B})}$, $M_{RL}=\sqrt{p_{B}(1-p_{A})}$ and $M_{RR}=\sqrt{(1-p_{B})(1-p_{A})}$. Regarding this state, the evolved state becomes 
 \begin{equation}
|\psi(\tau)\rangle_{AB} = e^{i\phi}(M_{LL}|LL\rangle +M_{LR} e^{i\Delta\phi_{LR}}|LR\rangle +M_{RL} e^{i\Delta\phi_{RL}}|RL\rangle +M_{RR} |RR\rangle), 
\end{equation}
and the density matrix becomes
 \begin{equation}\label{GES}
\rho_{AB}(\tau)=\left(\begin{array}{llll}
|M_{LL}|^{2} & M_{LL}M^{*}_{LR}e^{-i \Delta \phi_{L R}} & M_{LL}M^{*}_{RL} e^{-i \Delta \phi_{R L}}& M_{LL}M^{*}_{RR} \\
M_{LR}M^{*}_{LL}e^{i \Delta \phi_{L R}} & |M_{LR}|^{2} & M_{LR}M^{*}_{RL}e^{-i ( \Delta \phi_{R L}-\Delta \phi_{L R})} & M_{LR}M^{*}_{RR}e^{i \Delta \phi_{L R}}\\
M_{RL}M^{*}_{LL}e^{i \Delta \phi_{R L}} & M_{RL}M^{*}_{LR}e^{i (\Delta \phi_{R L}-\Delta \phi_{L R})} & |M_{RL}|^{2} & M_{RL}M^{*}_{RR} e^{i \Delta \phi_{R L}}\\
M_{RR}M^{*}_{LL} & M_{RR}M^{*}_{LR}e^{-i \Delta \phi_{L R}} & M_{RR}M^{*}_{RL}e^{-i \Delta \phi_{R L}} & |M_{RR}|^{2}
\end{array}\right).
\end{equation} 
One can obtain the negativity for this state as
\begin{equation}
\mathcal{N}(\rho_{AB})=4\sqrt{p_{A}p_{B}(1-p_{A})(1-p_{B})} \mid \sin(\frac{\Delta\phi_{LR} + \Delta\phi_{RL}}{2})\mid.
\end{equation}
As can be seen, if either subsystem is initially incoherent ($p_{A}$ or $p_{B}$ equal to zero or one), gravity cannot generate any entanglement (see Fig. \ref{fig6}). Another conclusion that can be drawn is that a maximally entangled final state is attainable only if both subsystems are initially maximally coherent ($p_{A}=p_{B} =\dfrac{1}{2}$). Figure \ref{fig7} displays the negativity as a function of $p_{A}$ and $p_{B}$. It is evident that changes in the initial coherence of the subsystems' states alter the amount of entanglement generated via gravity, and maximum entanglement is achieved when the subsystems' initial states are maximally coherent.\\
To extend the complementarity relations to arbitrary initial states, we consider the general evolved density matrix in Eq.\;(\ref{GES}). The corresponding reduced state for subsystem  $A$ is obtained as
 \begin{equation}
\rho_A=\left(\begin{array}{cc}
|M_{LL}|^{2}+|M_{LR}|^{2} & M_{LL}M^{*}_{RL}e^{-i \Delta \phi_{R L}}+M_{LR}M^{*}_{RR} e^{i \Delta \phi_{L R}} \\
M_{LL}M^{*}_{RL}e^{i \Delta \phi_{R L}}+M_{LR}M^{*}_{RR} e^{-i \Delta \phi_{L R}} & |M_{RL}|^{2}+|M_{RR}|^{2}
\end{array}\right).
\end{equation}
%%%%%%%%%%%%%%%%%%%%%%%%%%%%%%%%%
%%%%%%%%%%%%%%%%%%%%%%%%%%%%%%%%%%%%%%
%%%%%%%%%%%%%%%%%%%%%%%%%%%%%%%%%%%%%%%%%%
%%%%%%%%%%%%%%%%%%%%%%%%%%%%%%%%%%%%%%%%%%

%%%%%%%%%%%%%%%%%%%%%%%%%%%%%%%%%%%%%%
%%%%%%%%%%%%%%%%%%%%%%%%%%%%%%%%%%%%%%
%%%%%%%%%%%%%%%%%%%%%%%%%%%%%%%%%%%%%%%%%%
%%%%%%%%%%%%%%%%%%%%%%%%%%%%%%%%%%%%%%%%%%
From the above expression, a straightforward calculation yields the following relations for the generalized initial state:

 \begin{equation}
C_{l_{1}}(\rho_{A})^{2}+\mathcal{N}(\rho_{AB})^{2}\leqslant 1,
\end{equation}
 
\begin{equation}
C_{l_{1}}(\rho_{A})^{2}+\mathcal{C}(\rho_{AB})^{2}\leqslant 1,
\end{equation}
and
\begin{equation}
C_{r}(\rho_{A})+\mathcal{E}(\rho_{AB})\leqslant 1.
\end{equation}
These inequalities constitute the generalized complementarity relations between local coherence and entanglement for arbitrary initial product states. Crucially, these bounds are saturated—and reduce to the exact equalities in Eqs.\;(\ref{ComN}), (\ref{ComC}) and (\ref{ComR}) if and only if  the subsystems' initial states are maximally coherent ($p_{A}=p_{B}=\dfrac{1}{2}$). For any other initial coherence, the sum remains strictly less than one, demonstrating that the initial coherence acts as a finite resource that sets the maximum achievable entanglement. Moreover, these relations quantify the trade-off imposed by unitary gravitational evolution: coherence lost from a local subsystem is not destroyed but coherently transferred into shared entanglement, thereby conserving the total quantum resource.

%%%%%%%%%%%%%%
%%%%%%%%%%%%%%%
%%%%%%%%%%%%%%%
%%%%%%%%%%%%%%%%%%%%%%%%%%%%%%%%%%%
%%%%%%%%%%%%%%%%%%%%%%%%%%%%%%%%%%%%
\section{Conclusion}\label{conclusion}
In recent years, the study and search for signatures of the quantum nature of gravity have attracted the attention of many researchers. In this work, we have examined gravitational evolution from the perspective of quantum resource theory. We have shown that the gravitational interaction within the Bose \textit{et al.} model \cite{Bose} constitutes a unitary, coherence-preserving operation, which explains the conservation of total coherence during evolution. We have demonstrated that gravity acts as a channel  that coherently redistributes local quantum coherence, initially present in the spatial superposition of each mass, into shared bipartite entanglement. This conversion has been quantified by exact complementarity relations, which we have derived both for specific and general initial states. These relations establish that initial coherence is not merely beneficial but is a necessary resource for entanglement generation; its absence in any subsystem yields zero entanglement. Furthermore, the degree of initial coherence quantitatively bounds the maximum entanglement achievable, with maximal entanglement attainable only from initially maximal coherence. It is important to emphasize that the term "channel" in this context refers specifically to the unitary quantum evolution induced by the gravitational interaction, and not to a noisy or signaling channel in the sense of quantum communication theory. By clarifying this conversion mechanism, our analysis provides a refined, resource-theoretic foundation for interpreting future experiments. It highlights a critical experimental design criterion: successfully witnessing gravitationally induced entanglement necessitates preparing and maintaining a sufficient degree of quantum coherence in the test masses.

%=============================================================%
%=============================================================%
%=======================  References =========================%
%=============================================================%
%=============================================================%



%% References


%\bibliography{sample.bib}



%%%%%%%%%%%%%%%%%%%%%%%%%%%%%%%%%%%%%%%%%%%%%%%%%%%%%%%%%%%%%%%%%%%%%
%%%%%%%%%%%%%%%%%%%%%%%%%%%%%%%%%%%%%%%%%%%%%%%%%%%%%%%%%%%%%%%%%%%%%%%%%%%%%%%%%%%%%%%%%%%%%%%%%%%%%%%%%%%%%%%
%%%%%%%%%%%%%%%%%%%%%%%%%%%%%%%%%%%%%%%%%%%%%%%%%%%%%%%%%%%%%%%%%%%%

\end{document}